# Array of dipoles near a hyperbolic metamaterial: Evanescent-to-propagating Floquet wave transformation

Caner Guclu, Salvatore Campione, and Filippo Capolino,

Department of Electrical Engineering and Computer Science, University of California, Irvine, Irvine, 92697, CA, United States

f.capolino@uci.edu

*Abstract*—We investigate the capabilities of hyperbolic metamaterials (HMs) to couple near-fields (i.e., evanescent waves) emitted by a two-dimensional periodic array of electric dipoles to propagating waves. In particular, large order Floquet harmonics with transverse magnetic (TM) polarization, that would be evanescent in free space and therefore confined near the array surface, are transformed into propagating spectrum inside the HM, and thus carry power away. Because of this property, independent of the finite or infinite extent of the HM, the power generated by an array of elementary electric dipoles is strongly enhanced when the array is located near a HM surface and is mostly directed into the HM. In particular, the power coupled to the HM exhibits narrow frequency features that can be employed in detection applications. The results shown in this paper provide a clear signature on wave dynamics in HMs. A link between the results pertaining to the case of an isolated dipole on top of HM and the planar array is found convenient to explain both wave dynamics and spectral power distribution. The narrow frequency emission features appear in the array case only; they depend on its spatial periodicity and remarkable on the HM thickness.

PACS: 42.25.Fx 78.20.Ci 72.80.Tm 78.67.Pt

## I. INTRODUCTION

HYPERBOLIC metamaterials (HMs) are a subcategory of artificial uniaxial anisotropic materials that exhibit hyperbolic isofrequency wave-vector dispersion diagram [1-3].

HMs allow for engineering the spatial spectrum of propagating waves and thus power emission exploiting a wide propagating spectrum when compared to common dielectrics. This unusually wide spatial spectrum of power emission leads to novel phenomena such as the enhancement of the power scattered by nanospheres [4] or of the one emitted by imposed dipoles [4-6] located above HM surfaces. Furthermore, HMs are capable of absorbing (in the form of propagating waves) the power emitted by sources in their proximity. This, in turn, means that decay rate of emitters can be controlled without resorting to substrate's loss engineering. For this reason, HMs have been used to engineer the Purcell effect and emission decay rate, as well as the enhancement of spontaneous emission [7-14]. Moreover, the wide spatial spectrum supported by HMs leads to applications such as focusing with extreme subwavelength resolution and superlensing [15-22], as well as absorption and reflection control [23-25]. HMs have also been shown to exhibit negative refraction [26-29] and epsilon-near-zero capabilities [30, 31], and the latter could be for example employed to improve nonlinear processes. In [32] the formation of second harmonic double-resonance cones has been proven. Moreover, efficient second harmonic generation has been reported in [33] through the use of micrometer-thick slabs with hyperbolic permittivity tensor.

HMs can be fabricated at infrared and optical frequencies using metal-dielectric multilayers [9, 34], dielectric-semiconductor multilayers [35], graphene-dielectric multilayers [5, 28, 31, 36] or metallic wires embedded in dielectric substrates [25, 37]. In particular, the emergence of hyperbolic dispersion in multilayered HMs does not rely on any resonant behavior and thus occurs in a wide frequency band. A review of certain wave properties in HMs is reported in [4, 38-40]. We stress that practical HM realizations alter the ideal hyperbolic wavevector dispersion curve, and limit the propagating spectrum in contrast to what is predicted by the effective medium approximation (EMA) that does not introduce any limitation for the propagating spectrum in HMs [4, 6, 41].

In this study, we carry for the first time an analysis of electromagnetic waves generated by a two-dimensional (2D) periodic array of electric dipoles located above a HM. We first show how Floquet waves (FWs) emanating from such an array, that would be otherwise evanescent in free space, are instead transformed into propagating extraordinary waves inside a HM. We then investigate the enhancement of the power radiated by a 2D periodic array of impressed electric dipoles above a HM substrate, motivated by earlier work in which small scatterers on top of HMs (which can be modeled using single dipole approximation) or roughness on HM surfaces are shown to realize unprecedented absorption of plane waves [23, 25]. We show that most of the power generated by the 2D periodic array is directed towards the HM. We investigate how the array periodicity plays a critical role in the possibility to allocate wave propagation in the ideally indefinite spectral propagating channels of the HM.

We further show the effect of HM substrate's thickness on the properties mentioned above. For the first time, for the case of a 2D periodic array of sources on top of HM, we show the existence of very narrow-frequency emission peaks in both infinite and finite-thickness HMs, and then explain the theory thereof. Such peaks can be useful for sensing applications. Finally the discrete dipole approximation and its limitations are discussed over a representative example of array of rectangular current sheets (i.e., with limited physical domain).

## II. Coupling and Propagation of large-index Floquet Waves to a HM

The 2D periodic array of elementary electric dipoles is located above a HM at a distance $h$ from its surface as in Fig. 1(a). Electric dipolar sources in the array are located at $\mathbf{r}_{mn} = \mathbf{r}_{00} + ma\hat{\mathbf{x}} + nb\hat{\mathbf{y}}$ ($m, n = 0, \pm 1, \pm 2, ...$), where $a$ and $b$ are the periods along the $x$ and $y$ directions, respectively, and $\mathbf{r}_{00} = x_{00}\hat{\mathbf{x}} + y_{00}\hat{\mathbf{y}} + z_{00}\hat{\mathbf{z}}$ is the location of the reference dipole. In the following, the $00^{\text{th}}$ reference dipole is assumed to be located at $\mathbf{r}_{00} = \mathbf{0}$ (here we implicitly assume the time harmonic convention $e^{-i\omega t}$). Thus an electric dipole at $\mathbf{r}_{mn}$ has a dipole moment $\mathbf{p}_{mn} = \mathbf{p}_{00} \exp(i\mathbf{k}_t \cdot \mathbf{r}_{mn})$, where $\mathbf{k}_t = k_x\hat{\mathbf{x}} + k_y\hat{\mathbf{y}}$ is the wavevector defining the progressive phasing of the dipoles on the $x,y$ plane, and $\mathbf{p}_{00} = p_x\hat{\mathbf{x}} + p_y\hat{\mathbf{y}} + p_z\hat{\mathbf{z}}$ is the electric dipole moment of the $00^{\text{th}}$ reference dipole.

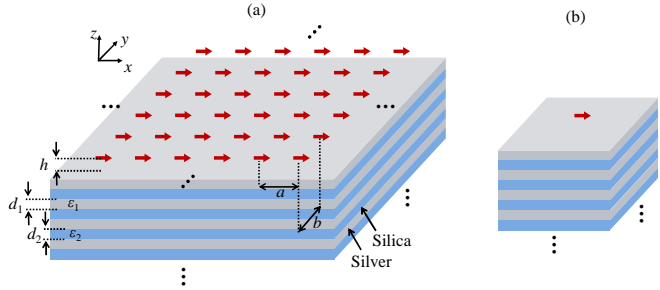

Fig. 1 (a) Schematic of an array of electric dipoles at a distance $h$ from the surface of a hyperbolic metamaterial. Example of HM made of a stack of dielectric and silver layers with thicknesses $d_1$ and $d_2$ and relative permittivities $\varepsilon_1$ and $\varepsilon_2$ . (b) Schematic of single dipole at a distance $h$ from the surface of a hyperbolic metamaterial in part (a).

We demonstrate here that near fields, in the form of FWs, emitted by a 2D periodic array of electric dipolar sources are coupled to propagating waves inside the HM substrate. It is well known that, in general depending on the array periodicity, only a certain (finite) number of FWs are propagating in a common dielectric, and thus carry power away from the array. The remaining FWs are in general evanescent and are confined mostly to the array plane forming the near field.

Consider for example the *direct* electric field produced by a 2D periodic array of elementary electric dipoles in free space, represented in terms of transverse-to-$z$ polarized electric (TE) and magnetic (TM) $pq$-indexed FWs as

$$\mathbf{E}(\mathbf{r}) = \sum_{p,q=-\infty}^{\infty} \left( \mathbf{E}_{pq}^{\text{TM}} + \mathbf{E}_{pq}^{\text{TE}} \right). \tag{1}$$

Each FW in free space is given by [42]

$$\mathbf{E}_{pq}^{\text{TM,TE}}(\mathbf{r}) = \frac{i}{2ab\varepsilon_0} \frac{e^{i\left(\mathbf{k}_{t,pq} \cdot \mathbf{r} + k_{z0,pq}|z|\right)}}{k_{z0,pq}} \mathbf{e}_{pq}^{\text{TM,TE}}, \tag{2}$$

with

$$
\begin{aligned}
\mathbf{e}_{pq}^{\text{TM}} &= \left\{ \begin{bmatrix} \dfrac{k_{z0,pq}^2}{k_{t,pq}^2}\mathbf{k}_{t,pq} \mp k_{z0,pq}\hat{\mathbf{z}} \end{bmatrix} \mathbf{k}_{t,pq} \right. \\
&\quad \left. + \begin{bmatrix} k_{t,pq}^2\hat{\mathbf{z}} \mp k_{z0,pq}\mathbf{k}_{t,pq} \end{bmatrix}\hat{\mathbf{z}} \right\} \cdot \mathbf{p}_{00}, \\
\mathbf{e}_{pq}^{\text{TE}} &= \frac{k_0^2}{k_{t,pq}^2}\left(\mathbf{k}_{t,pq} \times \hat{\mathbf{z}}\right)\left(\mathbf{k}_{t,pq} \times \hat{\mathbf{z}}\right) \cdot \mathbf{p}_{00},
\end{aligned}
\tag{3}
$$

where the minus (plus) sign in Eq. (3) is used when the observation point is above (below) the array plane. Longitudinal wavenumbers in free space, $k_{z0,pq}$, will be denoted with '0' in the subscript. The transverse wavevector of a $pq$-indexed FW is defined as

$$\mathbf{k}_{t,pq} = \mathbf{k}_{t,00} + \frac{2\pi p}{a}\hat{\mathbf{x}} + \frac{2\pi q}{b}\hat{\mathbf{y}}, \tag{4}$$

and $k_{t,pq}^2 = \mathbf{k}_{t,pq} \cdot \mathbf{k}_{t,pq}$. It is clear that the field in Eqs. (1)-(3) is intimately related to the transverse and longitudinal wavenumbers, $k_{t,pq}$ and $k_{z0,pq}$, respectively. In free space the relation of $k_{t,pq}$ and $k_{z0,pq}$ follows the free space isofrequency wavenumber dispersion equation, $k_{t,pq}^2 + k_{z0,pq}^2 = k_0^2$, with $k_0 = \omega/c$ the free space wavenumber, $\omega$ the angular frequency and $c$ the speed of light in vacuum. Therefore, for large $pq$-indexed FWs, the longitudinal wavenumber in free space $k_{z0,pq} = i\sqrt{k_{t,pq}^2 - k_0^2}$ is purely imaginary, with $\text{Im}\left(k_{z0,pq}\right) > 0$ to satisfy the radiation boundary condition at infinity (for "proper" waves, detailed information on proper and improper wave can be found in [42-45]). Such FWs are evanescent waves decaying exponentially away from the array plane (on the other hand, improper waves have opposite sign and grow exponentially toward infinity).

Consider now the scenario in Fig. 1(a), where the FWs generated by a 2D periodic array of electric dipoles with transverse wavenumber $\mathbf{k}_{t,pq}$ couple to a homogeneous HM characterized by the relative permittivity tensor

$$\underline{\varepsilon}_{\text{HM}} = \varepsilon_t \left( \hat{\mathbf{x}}\hat{\mathbf{x}} + \hat{\mathbf{y}}\hat{\mathbf{y}} \right) + \varepsilon_z \hat{\mathbf{z}}\hat{\mathbf{z}}, \tag{5}$$

with $\varepsilon_t \varepsilon_z < 0$ in an ideal lossless case, where $\varepsilon_t$ and $\varepsilon_z$ (both real numbers) represent the transverse and longitudinal entries of the diagonal relative permittivity tensor, respectively. Inside the HM, TM waves are extraordinary waves that satisfy the hyperbolic isofrequency dispersion relation [1, 2, 4] (assuming non-magnetic materials, thus relative permeability is taken as unity in the following)

$$\frac{k_{t,pq}^2}{\varepsilon_z} + \frac{\left(k_{z1,pq}^{\mathrm{TM}}\right)^2}{\varepsilon_t} = k_0^2, \qquad (6)$$

where $k_{t,pq}$ is matched to that of the FWs in free space above the HM (longitudinal wavenumbers in HM are denoted with '1' in the subscript). Isofrequency hyperbolic wavevector dispersion occurs when $\varepsilon_t \varepsilon_z < 0$, and we recall that two possible scenarios may occur, with either $\varepsilon_z < 0$ or $\varepsilon_t < 0$. The latter will be treated in Sec. IV pertaining to discussion and illustrative examples because it can be obtained at optical frequencies by simply stacking metal and dielectric layers of subwavelength thicknesses [4, 6, 29, 34].

Following Eq. (6), the $pq$-indexed TM FWs in the HM have longitudinal wavenumbers given by

$$k_{z1,pq}^{\mathrm{TM}} = \sqrt{\varepsilon_t\left(k_0^2 - \frac{1}{\varepsilon_z}k_{t,pq}^2\right)}, \qquad (7)$$

with sign of the square root chosen regarding the condition explained in the following. In general, assuming the presence of losses in the HM, the longitudinal wavenumber is complex and given as

$$k_{z1,pq}^{\mathrm{TM}} = \beta_{z1,pq}^{\mathrm{TM}} + i\alpha_{z1,pq}^{\mathrm{TM}}. \qquad (8)$$

A FW generated by the array at $z = 0$, with transverse wavenumber $k_{t,pq}^{\mathrm{TM}}$, assumes the wave propagator $\exp(ik_{z1,pq}^{\mathrm{TM}}|z|) = \exp(-ik_{z1,pq}^{\mathrm{TM}}z)$ along the $-z$ direction inside HM underneath. The condition $\alpha_{z1,pq}^{\mathrm{TM}} > 0$ is necessary to satisfy the boundary condition when $z$ tends to $-\infty$. Also, as explained in [4, 27, 29], waves in the HM are backward when $\varepsilon_t < 0$ and $\varepsilon_z > 0$ and hence characterized by $\beta_{z1,pq}^{\mathrm{TM}} < 0$, i.e., phase propagation occurs along $+z$ while power flows along $-z$. Indeed, as specified in [42-45] backward waves are characterized by a wavenumber that satisfies the relation $\beta_{z1,pq}^{\mathrm{TM}} \alpha_{z1,pq}^{\mathrm{TM}} < 0$.

Note that observing Eq. (7), when assuming absence of losses inside the HM, $k_{z1,pq}^{\mathrm{TM}}$ is *purely real* for *large pq* index values since the ratio $\varepsilon_t / \varepsilon_z$ is negative. This means that any TM FW with sufficiently large $pq$ order is able to propagate inside the HM with a real longitudinal wavenumber $k_{z1,pq}^{\mathrm{TM}}$. Note that FWs with small $pq$ indexes, in particular the fundamental one with $(p,q) = (0,0)$, may or may not be propagating. Low order FWs are propagating if $\varepsilon_t > 0$ whereas they are evanescent when $\varepsilon_t < 0$. Nevertheless, the most important phenomenon is that in theory an infinite number of FWs are able to propagate in an ideal HM with unlimited hyperbolic isofrequency wavevector dispersion curve. In practical realization of HMs however, the HM periodicity along a coordinate would restrict the range of FWs are able to propagate, as it will be briefly discussed in Sec. IV, and the power in FWs would strongly depend on the distance of the array from the HM and the presence of losses

Having clarified the propagation of TM and TE FWs inside a HM, we now analyze their excitation generated by a 2D periodic array of dipoles located at a distance $h$ below the array. We first assume that the HM is homogeneous with relative permittivity tensor as in Eq. (5) and it is semi-infinite. As previously described, the direct field produced by the array is represented as a sum of $pq$-indexed FWs (i.e., plane waves) as in Eqs. (1)-(3) for an array in free space. Each $pq$-indexed FW generated by the array and directed towards $-z$ is partly reflected at the free-space/HM interface, with TM/TE Fresnel reflection coefficient

$$\Gamma_{pq}^{\mathrm{TM,TE}} = \frac{Z_{\mathrm{HM}}^{\mathrm{TM,TE}}\left(\mathbf{k}_{t,pq}\right) - Z_0^{\mathrm{TM,TE}}\left(\mathbf{k}_{t,pq}\right)}{Z_{\mathrm{HM}}^{\mathrm{TM,TE}}\left(\mathbf{k}_{t,pq}\right) - Z_0^{\mathrm{TM,TE}}\left(\mathbf{k}_{t,pq}\right)}, \qquad (9)$$

conveniently given in terms of the characteristic wave impedances of free space

$$Z_0^{\mathrm{TM}}\left(\mathbf{k}_{t,pq}\right) = \frac{k_{z0,pq}}{\omega\varepsilon_0}, \qquad Z_0^{\mathrm{TE}}\left(\mathbf{k}_{t,pq}\right) = \frac{\omega\mu_0}{k_{z0,pq}}, \qquad (10)$$

and the characteristic wave impedances

$$Z_{\mathrm{HM}}^{\mathrm{TM}}\left(\mathbf{k}_{t,pq}\right) = \frac{k_{z1,pq}^{\mathrm{TM}}}{\omega\varepsilon_0\varepsilon_t}, \qquad Z_{\mathrm{HM}}^{\mathrm{TE}}\left(\mathbf{k}_{t,pq}\right) = \frac{\omega\mu_0}{k_{z1,pq}^{\mathrm{TE}}}, \qquad (11)$$

for the extraordinary (TM) and ordinary (TE) waves inside the HM, here assumed homogeneous. Note that $k_{z1,pq}^{\mathrm{TM}}$ for the extraordinary wave (TM) inside the HM is evaluated as in Eq. (7), whereas $k_{z1,pq}^{\mathrm{TE}}$ for the ordinary (TE) wave is evaluated by the ordinary wave dispersion relation $k_{t,pq}^2 + \left(k_{z1,pq}^{\mathrm{TE}}\right)^2 = \varepsilon_t k_0^2$, and thus

$$k_{z1,pq}^{\mathrm{TE}} = \sqrt{\varepsilon_t k_0^2 - k_{t,pq}^2} \qquad (12)$$

As discussed above, for TM waves with large $pq$ indexes, $k_{z1,pq}^{\mathrm{TM}}$ is real, assuming a lossless HM, and thus also $Z_{\mathrm{HM}}^{\mathrm{TM}}\left(\mathbf{k}_{t,pq}\right)$ is real. Furthermore, in the case of a HM with $\varepsilon_t < 0$ considered here we conclude after observing Eq. (11) that $Z_{\mathrm{HM}}^{\mathrm{TM}}\left(\mathbf{k}_{t,pq}\right)$ is real positive because a TM wave in the HM is backward (i.e., $\beta_{z1,pq}^{TM} < 0$) as discussed previously. Therefore, in a HM, including the occurrence of losses, the characteristic wave impedance $Z_{HM}^{TM}\left(\mathbf{k}_{t,pq}\right)$ as in Eq. (11) has a positive real part which is associated to power flowing in the HM in the $-z$ direction. [Here, the general expression of a characteristic wave impedance $Z\left(\mathbf{k}_{t,pq}\right)$ is defined assuming that the real part of the $z$-directed (in the positive or negative $z$ direction) Poynting vector is given by $S\left(\mathbf{k}_{t,pq}\right) = \frac{1}{2}\mathrm{Re}\left[\left|\mathbf{E}_{t,pq}\right|^2 / Z^*\left(\mathbf{k}_{t,pq}\right)\right]$, where $\mathbf{E}_{t,pq}$ is the transverse field of the harmonic, yielding always positive $S\left(\mathbf{k}_{t,pq}\right)$ for waves carrying power away from the array (note that a '*' denotes complex conjugation).]

Each FW generated by the array above the HM is partly transmitted into the HM with transmission coefficient $T_{pq}^{TM,TE} = 1 + \Gamma_{pq}^{TM,TE}$. (We recall that all reflection and transmission coefficients $\Gamma_{pq}^{TM,TE}$ and $T_{pq}^{TM,TE}$ are defined with respect to transverse electric fields [46].) Therefore the transverse field at any location $\mathbf{r}$ above the 2D periodic array of dipoles (i.e., $z > 0$) is represented as

$$\mathbf{E}_t\left(\mathbf{r}\right) = \sum_{p,q=-\infty}^{\infty} \left[ \begin{array}{l} \left(\mathbf{E}_{t,pq}^{TM+} + \mathbf{E}_{t,pq}^{TM-}\Gamma_{pq}^{TM}e^{i2k_{z0,pq}h}\right) + \\ + \left(\mathbf{E}_{t,pq}^{TE+} + \mathbf{E}_{t,pq}^{TE-}\Gamma_{pq}^{TE}e^{i2k_{z0,pq}h}\right) \end{array} \right] e^{ik_{z0,pq}z}, \tag{13}$$

where the subscript $t$ denotes the transverse component of the electric field $\mathbf{E}_{pq}^{TM,TE}$ in Eq. (2) and +/- superscripts denote that the respective quantity is evaluated at the limit $z \to 0^+$ and $z \to 0^-$, respectively [the array plane is assumed at $z = 0$]. Note that the transverse component of $\mathbf{E}_{pq}^{TE}$ is always continuous across the array plane, i.e. $\mathbf{E}_{t,pq}^{TE+} = \mathbf{E}_{t,pq}^{TE-}$, and TE FWs are only emitted by transverse-to-$z$ dipole components. On the other hand, the transverse component of $\mathbf{E}_{pq}^{TM}$ should be treated carefully. For example when $\mathbf{p}_{00}$ is transverse to $z$, the transverse component of $\mathbf{E}_{pq}^{TM}$ is continuous across the array plane, i.e. $\mathbf{E}_{t,pq}^{TM+} = \mathbf{E}_{t,pq}^{TM-}$; however when $\mathbf{p}_{00}$ is along $z$, the transverse component of $\mathbf{E}_{pq}^{TM}$ follows the relation $\mathbf{E}_{t,pq}^{TM+} = -\mathbf{E}_{t,pq}^{TM-}$ [dictated by Eq. (3)].

At any location $\mathbf{r}$ below the array of dipoles and above the HM (i.e., $-h < z < 0$) the transverse electric field is

$$\mathbf{E}_t\left(\mathbf{r}\right) = \sum_{p,q=-\infty}^{\infty} \left[ \begin{array}{l} \mathbf{E}_{t,pq}^{TM-}\left(e^{-ik_{z0,pq}z} + \Gamma_{pq}^{TM}e^{ik_{z0,pq}(2h+z)}\right) + \\ + \mathbf{E}_{t,pq}^{TE-}\left(e^{-ik_{z0,pq}z} + \Gamma_{pq}^{TE}e^{ik_{z0,pq}(2h+z)}\right) \end{array} \right]. \tag{14}$$

The transverse field transmitted to a homogeneous HM (i.e., at any location $\mathbf{r}$ belonging to the HM, with $z < -h$) is represented as

$$\mathbf{E}_t\left(\mathbf{r}\right) = \sum_{p,q=-\infty}^{\infty} \left[ \begin{array}{l} \mathbf{E}_{t,pq}^{TM-}T_{pq}^{TM}e^{-ik_{z1,pq}^{TM}(z+h)} + \\ + \mathbf{E}_{t,pq}^{TE-}T_{pq}^{TE}e^{-ik_{z1,pq}^{TE}(z+h)} \end{array} \right] e^{ik_{z0,pq}h}. \tag{15}$$

Looking at Eq. (15), it is clear that the distance $h$ plays a fundamental role in determining the spectrum of evanescent waves in free space that can be coupled to propagating waves in the HM. In particular, when the array is located at a certain distance $h$ from the HM surface, waves will decay with the propagator $e^{ik_{z0,pq}h}$, resulting in a decay factor $e^{-\mathrm{Im}(k_{z0,pq})h}$, in free space, and this will prevent high $pq$-indexed FWs from transferring power to the HM underneath.

## III. POWER GENERATED BY A 2D PERIODIC ARRAY OF ELECTRIC DIPOLES ABOVE A HM

The real power density (the real part of the Poynting vector) of each $pq$-th FW in $(+z)$ upward direction above the array is expressed as

$$S_{up}\left(\mathbf{k}_{t,pq}\right) = \frac{1}{2}\mathrm{Re}\left[\frac{\left|\mathbf{E}_{t,pq}^+ + \mathbf{E}_{t,pq}^-\Gamma_{pq}e^{i2k_{z0,pq}h}\right|^2}{Z_0^*\left(\mathbf{k}_{t,pq}\right)}\right], \tag{16}$$

where $Z_0\left(\mathbf{k}_{t,pq}\right)$ is the impedance looking upward.

Analogously, the real power density of each $pq$-th FW evaluated at the HM interface in the downward direction, and thus entering the HM, is given by

$$S_{down}\left(\mathbf{k}_{t,pq}\right) = \frac{1}{2}\mathrm{Re}\left[\frac{\left|\mathbf{E}_{t,pq}^-T_{pq}e^{ik_{z0,pq}h}\right|^2}{Z_{HM}^*\left(\mathbf{k}_{t,pq}\right)}\right] \tag{17}$$

where $Z_{HM}\left(\mathbf{k}_{t,pq}\right)$ is the impedance looking downward. The superscript TM/TE is omitted in Eqs. (16) and (17) since both expressions are valid for both TM and TE waves, assuming all quantities are evaluated accordingly. The superscripts +/- follow the same convention introduced in Sec. II.

In Eqs. (16) and (17) numerators are real valued; in contrast, denominators may be complex valued, depending on the medium where each FW is propagating into and the transverse wavevector $\mathbf{k}_{t,pq}$. In particular, $Z_0\left(\mathbf{k}_{t,pq}\right)$ is real only when $k_{t,pq}^2 < k_0^2$ and purely imaginary otherwise (for both TM and TE waves). In other words, only low $pq$-indexed FWs, with real $k_{z0,pq}$, carry power away from the array in the upward direction. Similarly, $Z_{HM}^{TE}\left(\mathbf{k}_{t,pq}\right)$ is (assuming lossless HM) either purely imaginary for any $k_{t,pq}$ (when $\varepsilon_t < 0$ and $\varepsilon_z > 0$), or purely real only for the spectrum $k_{t,pq}^2 < \varepsilon_t k_0^2$ (when $\varepsilon_t > 0$ and $\varepsilon_z < 0$). However for extraordinary waves with TM polarization inside the HM, the situation is rather different. The term $Z_{HM}^{TM}\left(\mathbf{k}_{t,pq}\right)$ has a real part (purely real for a lossless HM as discussed in Sec. II) for large $pq$-indexed FWs because $k_{z1,pq}^{TM}$ has a real part (purely real for a lossless HM) for large $pq$ indexes as described in Sec. II. Therefore, the total power coupled from the array in free space to the HM underneath is determined by the power carried by a very large number (infinite for an ideal lossless HM) of propagating FWs hosted by HM (however, note that the power coupled to FW with large $pq$ is also limited by the decay along the distance $h$ as described in Sec. II). Indeed, as it will be discussed in later sections, practical HM implementations based on periodic arrangement of layers or other configurations (e.g., wire medium) limit the maximum $pq$ indexes of FWs that can propagate in HM, thus limiting the maximum amount of power coupled to HM. Equation (17) is valid for a homogeneous HM (that includes losses) and it is generalized to the case of multilayer HM by substituting $T_{pq}$ with $1 + \Gamma_{pq}$, and $Z_{HM}\left(\mathbf{k}_{t,pq}\right)$ with the impedance at the multilayer HM interface, in the downward direction, $Z_{down}\left(\mathbf{k}_{t,pq}\right)$, as explained in [5]. The field inside the HM multilayer can be evaluated via transfer matrix method.

We now investigate in detail the power coupled to TM and TE plane wave spectra, based on [1, 47, 48], adapted to periodic structures, as in [49], for modeling the power emitted by a 2D periodic array of impressed (transverse or vertical) electric dipoles located slightly above an infinitely extended HM as in Fig. 1(a). Then we establish the relation between the plane wave spectra emanating from an array and a single dipole [as in Fig. 1(b)] above the same HM. Note that this TL formalism represents the solution of Maxwell's equations in the analyzed environment, i.e., it is an exact representation [1, 47, 48].

We are interested in the power emitted in $+z$ and $-z$ directions, in the following denoted by superscripts "up" and "down", respectively. The power emitted in a *unit cell* directed up/down for an array of *transverse dipoles* with $\mathbf{p}_{00} = p_x\hat{\mathbf{x}} + p_y\hat{\mathbf{y}}$ is given as a sum of TM and TE contributions as

$$P_{up/down} = \frac{\omega^2}{2ab} \sum_{p,q=-\infty}^{\infty} \begin{bmatrix} U_{up/down}^{TM}(\mathbf{k}_{t,pq}) + \\ + U_{up/down}^{TE}(\mathbf{k}_{t,pq}) \end{bmatrix}. \quad (18)$$

expressed in [W]. The TM and TE power spectra in Eq. (18) (normalized by angular frequency squared, $\omega^2$) directed toward the $+/-z$ directions are

$$U_{up/down}^{TM}(\mathbf{k}_{t,pq}) = \frac{\left|\mathbf{p}_{00} \cdot \mathbf{k}_{t,pq}\right|^2}{\left|\mathbf{k}_{t,pq}\right|^2} \\ \times \frac{\mathrm{Re}\left(Y_{up/down}^{TM*}(\mathbf{k}_{t,pq})\right)}{\left|Y_{up}^{TM}(\mathbf{k}_{t,pq}) + Y_{down}^{TM}(\mathbf{k}_{t,pq})\right|^2}, \quad (19)$$

$$U_{up/down}^{TE}(\mathbf{k}_{t,pq}) = \frac{\left|\mathbf{p}_{00} \cdot \left(\hat{\mathbf{z}} \times \mathbf{k}_{t,pq}\right)\right|^2}{\left|\mathbf{k}_{t,pq}\right|^2} \\ \times \frac{\mathrm{Re}\left(Y_{up/down}^{TE*}(\mathbf{k}_{t,pq})\right)}{\left|Y_{up}^{TE}(\mathbf{k}_{t,pq}) + Y_{down}^{TE}(\mathbf{k}_{t,pq})\right|^2}. \quad (20)$$

The power emitted in a unit cell in an array of *vertical dipoles* with $\mathbf{p}_{00} = p_z\hat{\mathbf{z}}$ is instead

$$P_{up/down} = \frac{\omega^2}{2ab} \sum_{p=-\infty}^{\infty} \sum_{q=-\infty}^{\infty} W_{up/down}^{TM}(\mathbf{k}_{t,pq}), \quad (21)$$

where the power spectrum (normalized by angular frequency squared, $\omega^2$), which has contribution only from TM waves, is

$$W_{up/down}^{TM}(\mathbf{k}_{t,pq}) = \frac{\left|\mathbf{p}_{00} \cdot \hat{\mathbf{z}}\right|^2}{\omega^2 \left|\varepsilon\right|^2} \left|\mathbf{k}_{t,pq}\right|^2 \\ \times \frac{\mathrm{Re}\left(Z_{up/down}^{TM}(\mathbf{k}_{t,pq})\right)}{\left|Z_{up}^{TM}(\mathbf{k}_{t,pq}) + Z_{down}^{TM}(\mathbf{k}_{t,pq})\right|^2}. \quad (22)$$

Here $Y_{up/down}^{TM/TE}$ and $Z_{up/down}^{TM/TE}$ are the equivalent admittances and impedances experienced by TM/TE waves, respectively, evaluated at the array plane and computed via transfer matrix method, for FWs emanating from a 2D periodic array of electric dipoles as in Fig. 1(a) traveling upward and downward, respectively, as a function of the transverse wavevector $\mathbf{k}_{t,pq}$.

In the following we establish a relation between the power expressions related to the 2D periodic array of electric dipoles to that of a single electric dipole on top of HM as in Fig. 1(b), analyzed in [4] for example. According to the expressions presented in [4] for the *single dipolar source* case, the upward and downward directed power emitted by a *transverse dipole* are

$$P_{\text{up/down}} = \frac{\omega^2}{8\pi^2} \iint \begin{bmatrix} U_{\text{up/down}}^{\text{TM}}(\mathbf{k}_t) + \\ + U_{\text{up/down}}^{\text{TE}}(\mathbf{k}_t) \end{bmatrix} dk_x dk_y. \quad (23)$$

The power emitted by a *vertical dipole* is

$$P_{\text{up/down}} = \frac{\omega^2}{8\pi^2} \iint W_{\text{up/down}}^{\text{TM}}(\mathbf{k}_t) dk_x dk_y. \quad (24)$$

Note that the functions $U$ and $W$ within the summations in Eqs. (18) and (22), for the periodic array case, are the same functions as the integrands in Eqs. (23) and (24), for the single dipole. In other words, the radiation from the array corresponds to a spectral sampling of the continuous spectrum of waves emanating from a single source. This fact emphasizes the importance of array periodicity to control effectively the sampling in the band of propagating FWs in HM, as it will be pointed out in what follows.

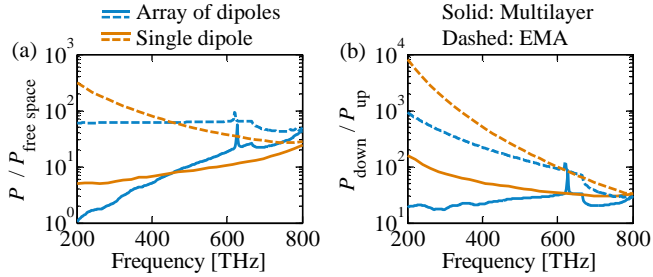

Fig. 2. (a) Enhancement of the power $P = P_{\text{up}} + P_{\text{down}}$ emitted by an array of dipoles ($a = b = 300$ nm) with a HM underneath with respect to that emitted in free space. Dipoles are polarized along $x$, with $\mathbf{k}_{t,00} = 0.5k_0\hat{\mathbf{x}}$. For comparison, also the power enhancement pertaining to a single dipole over the HM is provided. (b) Ratio of power emitted by the array and a single dipole towards HM (down) and towards the upper homogeneous isotropic space (up) versus frequency. Dashed lines are obtained for a homogeneous HM (via EMA) made of silver and silica layers with equal thicknesses $d_1 = d_2 = 10$ nm whereas solid lines are obtained using a rigorous multilayer Green's function implementation for the same HM. Silver permittivity $\varepsilon_2$ is from [50] and dielectric relative permittivity $\varepsilon_1$ is equal to 2.2. Source distance from the HM is assumed as $h = 10$ nm.

Before going into the details of the wave dynamics of the system under analysis, we provide two preliminary examples showing how the HM impacts the emitted power and the ratio of the power directed toward the HM for an array of dipoles against a single dipole for an infinitely extended HM substrate. Here we utilize a semi-infinite practical HM implementation consisting of a stacked bi-layers made of silver (whose dielectric relative permittivity function $\varepsilon_2$ is taken from the experimental results including losses in [50]) and silica ($\varepsilon_1 = 2.2$) layers with equal thicknesses $d_1 = d_2 = 10$ nm. The power evaluations are carried out using HMs modeled with two methods: the effective medium approximation (EMA) as given in [51] where

$$\varepsilon_t = \frac{\varepsilon_1 d_1 + \varepsilon_2 d_2}{d_1 + d_2} \qquad \varepsilon_z^{-1} = \frac{\varepsilon_1^{-1} d_1 + \varepsilon_2^{-1} d_2}{d_1 + d_2}. \quad (25)$$

and the more accurate multilayer Bloch analysis [46] based on the transfer matrix method in evaluation of the impedances $Z_{\text{down}}(\mathbf{k}_{t,pq})$.

In Fig. 2(a), we report the enhancement of the total power $P = P_{\text{up}} + P_{\text{down}}$ emitted by an array of dipoles with respect to the total power emitted by the array in free space. The power enhancement for an array of dipoles in general increases as the frequency is increased. As a comparison we also provide the enhancement for a single dipole which exhibits less dependence on frequency and the enhancement is larger than the one in the array case at lower frequencies whereas at higher frequencies, the array case shows more enhancement due to periodicity of the sources as will be shown in Sec. IV. Moreover the EMA vastly overestimates the power enhancement at lower frequencies for both cases; this is a well-known shortcoming [4, 6, 41] and we will recall the reasons behind it in the next section. When using the rigorous multilayer Bloch theory (it is exact when assuming ideally smooth surface boundaries), the enhancement for the array case is between 10 to 30 folds for frequencies between 500 and 800 THz, where EMA yields at least one order more enhancement at low frequency. In Fig. 2(b), we report the ratio $P_{\text{down}} / P_{\text{up}}$ for the cases in Fig. 2(a). When using the Bloch model, we observe that the array has $P_{\text{down}} / P_{\text{up}}$ ratio between 20 and 30 over the whole frequency range whereas the ratio is higher for the single dipole case especially at the lower frequencies. The overestimation of $P_{\text{down}} / P_{\text{up}}$ by the homogenized HM model (EMA) is also observed, particularly at low frequency. We further show in Fig. 3 the ratios $P / P_{\text{free space}}$ and $P_{\text{down}} / P_{\text{up}}$ by changing the period of the array of dipoles along the $x$ and $y$ axes, calculated assuming a multilayer HM. We observe that as the period increases both $P / P_{\text{free space}}$ and $P_{\text{down}} / P_{\text{up}}$ increase. A period of $a = b = 300$ nm (half a wavelength at 500 THz) leads to an enhancement almost more than 10 folds above 500 THz and it increases at higher frequencies; the power is mostly directed to the HM over the whole frequency range shown here (300-800 THz), especially for larger period.

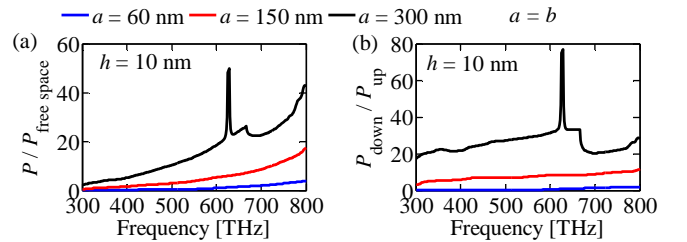

Fig. 3. (a) Emitted power $P = P_{\text{up}} + P_{\text{down}}$ by an array of dipoles (polarized along $x$ and with $\mathbf{k}_{t,00} = 0.5k_0\hat{\mathbf{x}}$) and (b) the ratio of power emitted by the array towards HM and towards free space versus frequency with respect to different array periods $a = b$. The HM is as in Fig. 2. This result is calculated assuming a multilayer HM.

It is important to note for the array case the occurrence of the peaks at about 627 THz for $a = b = 300$ nm case only; we attribute this peak to the surface plasmon polariton (SPP) at the interface of free space and HM for which the modal wavenumber elaboration is provided in the Appendix. Moreover similar narrow-frequency peaks are also present at many more frequencies when considering finite-thickness HMs. These peaks are particular to the array of dipoles, and are not observed for the single dipole emission on top of HM. The reasons of these peaks will be explained in Sec. V. In the next section it is proven that we can achieve enhanced, coupled power toward HM by proper design of the array on top of HM.

## IV. DISCUSSION AND ILLUSTRATIVE EXAMPLES

We consider here the same HM as introduced in the previous section assumed to have semi-infinite extent in $z$, and investigate the dynamics of the FWs emitted by an array of dipoles. The array of electric dipoles oscillating at 650 THz (i.e., a free-space wavelength of $\lambda_0 = 462$ nm) is assumed to be located in free space at a distance $h = 10$ nm from the HM (subwavelength proximity).

We first recall the case for a single emitting dipole as in Fig. 1(b) that will be helpful in subsequent analyses. The total power spectra (in logarithmic scale) are given in Fig. 4 versus $k_x$ and $k_y$ for both a transverse unit electric dipole located as in Fig. 1(b) with $\mathbf{p}_{00} = (\hat{\mathbf{x}} + \hat{\mathbf{y}})//\sqrt{2}$ Cm [evaluated as $U_{\text{up}}^{\text{TM}} + U_{\text{down}}^{\text{TM}} + U_{\text{up}}^{\text{TE}} + U_{\text{down}}^{\text{TE}}$, terms taken from Eqs. (19) and (20)] and a vertical unit electric dipole $\mathbf{p}_{00} = \hat{\mathbf{z}}$ [evaluated as $W_{\text{up}}^{\text{TM}} + W_{\text{down}}^{\text{TM}}$, terms taken from Eq. (22)].

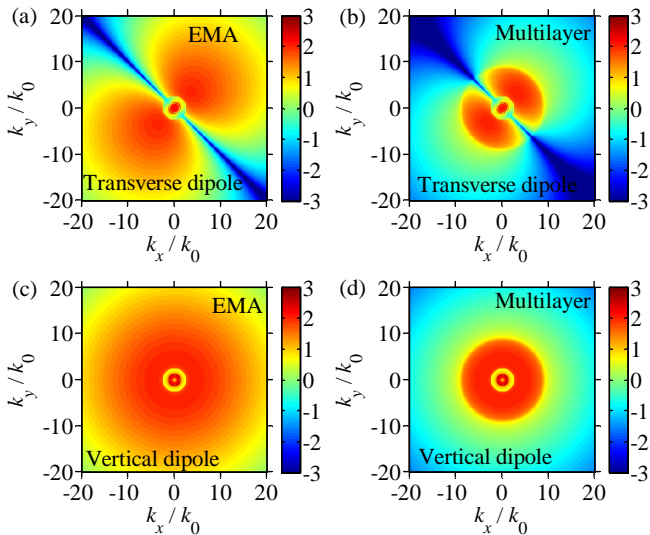

Fig. 4. Total spectral power versus $k_x$ and $k_y$ (a, b) $\log_{10}\left[\left(U_{\text{up}}^{\text{TM}} + U_{\text{down}}^{\text{TM}} + U_{\text{up}}^{\text{TE}} + U_{\text{down}}^{\text{TE}}\right)/\left(\text{Wm}^2\text{s}^2\right)\right]$ emitted by the unit

transverse electric dipole $\mathbf{p}_{00} = (\hat{\mathbf{x}} + \hat{\mathbf{y}})//\sqrt{2}$ Cm, (c, d) $\log_{10}\left[\left(W_{\text{up}}^{\text{TM}} + W_{\text{down}}^{\text{TM}}\right)/\left(\text{Wm}^2\text{s}^2\right)\right]$ emitted by the unit vertical dipole $\mathbf{p}_{00} = \hat{\mathbf{z}}$. In (a) and (c) EMA is used in HM modeling whereas in (b) and (d) multilayer HM is assumed, at 650 THz.

We first note in Fig. 4 the strong power spectrum for large transverse wavevectors. We also observe a strong dependence on the transverse wavevector direction for the transverse dipole, and no dependence for the vertical one due to symmetry reasons. Fundamentally, for metal-dielectric HMs the spectra provided in Fig. 4 is a wide-frequency phenomenon that does not rely on resonant characteristics, and it is provided for a representative frequency.

Moreover, we note the presence of a much wider spectrum of waves carrying power in the HM when looking at EMA results than multilayer ones. This result clearly explains that EMA overestimates power quantities, and is in agreement with previous investigations [4, 6, 41]. Since Bloch analysis models accurately the HM dispersion properties and EMA overestimates features for high spectral regions we will use Bloch multilayer modeling from this point on.

We then turn our attention to understanding which waves carry most of the power. To do so, we plot in Fig. 5 the spectral power $U_{\text{up/down}}^{\text{TM/TE}}$ in Eqs. (19) and (20) coupled to TM and TE waves toward both upper and bottom half spaces at 650 THz for the transverse dipole case in Fig. 4(b). It is clear that the power is mostly emitted in TM spectrum in $-z$ direction, i.e., towards the HM. A similar situation is encountered when analyzing the spectral power $W_{\text{up/down}}$ in Eq. (22) for a vertical dipole (not shown for brevity).

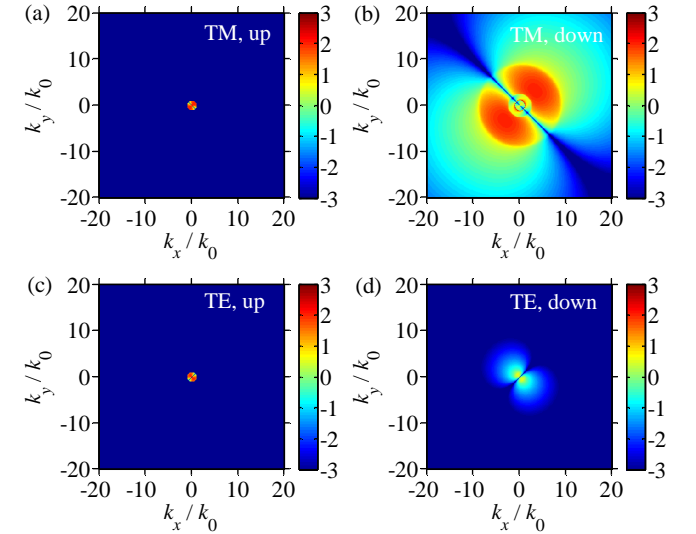

Fig. 5. Spectral power $\log_{10}\left[U_{\text{up/down}}^{\text{TM/TE}}/\left(\text{Wm}^2\text{s}^2\right)\right]$ versus $k_x$ and $k_y$ emitted by the unit transverse dipole $\mathbf{p}_{00} = (\hat{\mathbf{x}} + \hat{\mathbf{y}})//\sqrt{2}$ Cm in (a) $U_{\text{up}}^{\text{TM}}$: TM polarization and $+z$ direction, (b) $U_{\text{down}}^{\text{TM}}$: TM polarization and $-z$ direction, (c) $U_{\text{up}}^{\text{TE}}$: TE polarization and $+z$ direction, and (d) $U_{\text{down}}^{\text{TE}}$: TE polarization and $-z$ direction. This result is calculated assuming a multilayer HM.

The information in Fig. 4 and Fig. 5 will be now used to study the case of a 2D periodic array of electric dipoles on top of HM as in Fig. 1(a). Indeed, the spectral power quantities discussed in Figs. 4 and 5 are subject to sampling in the case of an array of dipoles as mentioned in Sec. II. Here we assume the array's progressive phasing is governed by $\mathbf{k}_{t,00} = 0.5k_0\hat{\mathbf{x}}$ where $k_0$ is the free-space wavenumber, and we investigate the spectral power for three sets of periods (assuming square lattice): 150, 300, and 600 nm. On the left panels of Fig. 6, the power in FW harmonics (normalized by angular frequency squared $\omega^2$) is reported versus FW indices $p$ and $q$. The larger the period, the more is the number of propagating FWs carrying power away as discussed in Sec. II. On the right panels of Fig. 6, we show the spectral power map pertaining to a single transverse dipole on top of HM where we superimpose the sampling points due to array periodicity (white circles) [the sampling procedure was mentioned in Sec. III]. It is clear that as the period increases the spectral power in the case of the array resembles that of the single dipole. This result indeed demonstrates that, for increasing periods, dipoles in the array experience less and less coupling between each other, therefore the array response tends to be very similar to an isolated dipole. Moreover, it is evident that periodicity can be optimized to couple most of the power to propagating extraordinary waves in HM.

versus $k_x$ and $k_y$ emitted by the unit transverse dipole $\mathbf{p}_{00} = (\hat{\mathbf{x}}+\hat{\mathbf{y}})/\sqrt{2}$ Cm at 650 THz where the white circles denote the Floquet harmonic sampling locations on $k_x$ - $k_y$ plane in the array case for various array periods as in (a, c, e). This result is calculated assuming a multilayer HM.

Indeed, the impact of the period, thus the sampling of the spatial spectral power, manifests itself in the enhancement and upward/downward redistribution of the emitted power. In order to achieve enhancement of emitted power with respect to free space, $P/P_{\text{free space}}$, one needs to sample at as many points as possible inside the propagating spectrum of HM with high $k_x$ and $k_y$. The increase in the period realizes this with a large ratio of $P_{\text{down}}/P_{\text{up}}$ as well. On the other hand if an array of dipolar sources were to represent induced dipoles modeling polarized scatterers, the fraction of the scattered power to the power impinging on a unit cell would decrease with increasing period, thus this decrease would undesirably downplay the coupling to the propagating spectrum in HM. A critical balance must be determined for this situation and will be studied in the future.

## V. Floquet waves Coupled to Modes in HM: Array Over Finite-thickness HM Substrates

We analyze the effect of finite HM thickness as in Fig. 7 on the power emission enhancement and redistribution. The impact of the number of the bi-layer unit cells $N$ on the spectral power and then on the enhancement and redistribution of the emitted power is demonstrated.

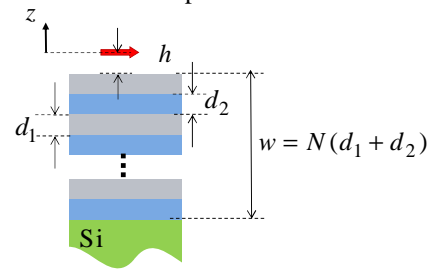

Fig. 7. HM substrate with finite thickness where $N$ is the number of metal-dielectric bi-layers. This substrate configuration is investigated for both single dipolar source and an array of sources.

The spectral power density sampling scheme explained in the previous section stresses the relationship between the power emitted into FWs and the spectral power emitted by a single dipole. Therefore the study of spectral power due to a single source over a finite-thickness HM is fundamental to characterize the emission from an array in the same setup. We start by showing in Fig. 8 how the number of bi-layers affects the spectral power distribution, generated by a single dipole polarized along $x$, for varying number of bi-layers $N = 1, 5, 10$ and $N \to \infty$, using the same HM composition as in the previous examples. We first observe that the power spectrum is strong over a wide wavevector space, a sign that large wavenumber waves are actually able to transport energy

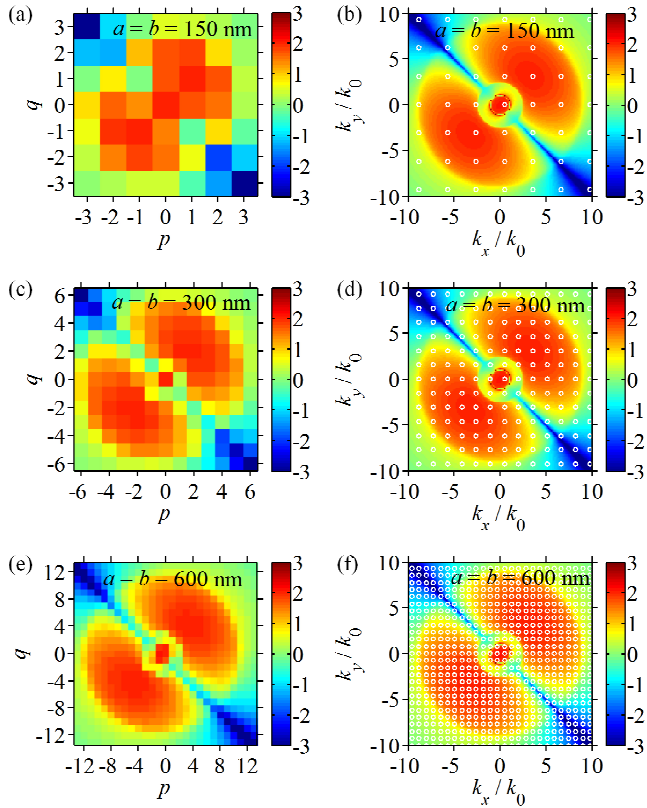

Fig. 6. (a, c, e) The spectral power of the Floquet harmonics versus the indices $p$ and $q$ for the array periods $a = b =$ 150, 300, 600 nm; (b, d, f) total spectral power ($\log_{10}\left[\left(U_{\text{up}}^{\text{TM}} + U_{\text{down}}^{\text{TM}} + U_{\text{up}}^{\text{TE}} + U_{\text{down}}^{\text{TE}}\right)/\left(\text{Wm}^2\text{s}^2\right)\right]$)

away from the array. Furthermore, we observe circular "belts" of spectral peaks in high $k_x$ and $k_y$ regions. The number of peaks depends on the number of metal-dielectric layers, in agreement with the bulk plasmon modes reported in [40]. As the number of layers $N$ tends to infinity in Fig. 8 (d), the field inside the HM is composed of a more uniform spectrum of propagating waves guided by the HM substrate. Next we stress the presence of the peak representing the wavenumber spectrum coupled to the SPP mode on the interface of free space and HM whose existence and wavenumber are determined in the Appendix for the case of homogeneous HM. The power coupled to this mode is in the spectrum slightly larger than $|\mathbf{k}_t| = k_0$ circle and it is clearly visible in the close-up view in Fig. 8(d) which is present in all cases reported in Fig. 8, but not so well defined for $N$=1. We stress that this innermost circular peak of the spectrum in the vicinity of circle with the radius $k_0$, remains a distinct spectral feature as $N$ increases, even when the aforementioned spectrum becomes uniform for $N \to \infty$. This distinct spectral feature will, of course, affect the total emitted power by an array when some spectral sampling point lays on it, as discussed next.

harmonics, which depends on the frequency, results in narrow frequency peaks of the total emitted power and the power in the downward direction as in Fig. 9(a) and (c). This will be further justified by the discussion relative to Fig. 10. Moreover for array of dipoles, we observe that as $N$ increases, $P / P_{\text{free space}}$ and $P_{\text{down}} / P_{\text{up}}$ become more stable versus frequency whereas for $N = 1$ the peaks are sharper and $P_{\text{down}} / P_{\text{up}}$ becomes lower than other cases at lower frequencies. This is due to several FW harmonics sampling the very sharp circular peak in spectral power [given in Fig. 8(a) for a certain frequency] when varying the frequency. In single dipole case, it is clear that even $N = 5$ is enough to emulate the effect of $N \to \infty$. Therefore, having greater $N$ causes larger values of $P / P_{\text{free space}}$ and $P_{\text{down}} / P_{\text{up}}$ at low frequency with respect to the $N = 1$ case. For the array of dipoles, when $N \to \infty$, we still observe a peak at 627.4 THz, absent in the single dipole case. This particular peak is due to the innermost peak circle in Fig. 8 [a close up view is provided in Fig. 8(d) representative for other cases as well] that corresponds to the SPP, briefly described in the Appendix.

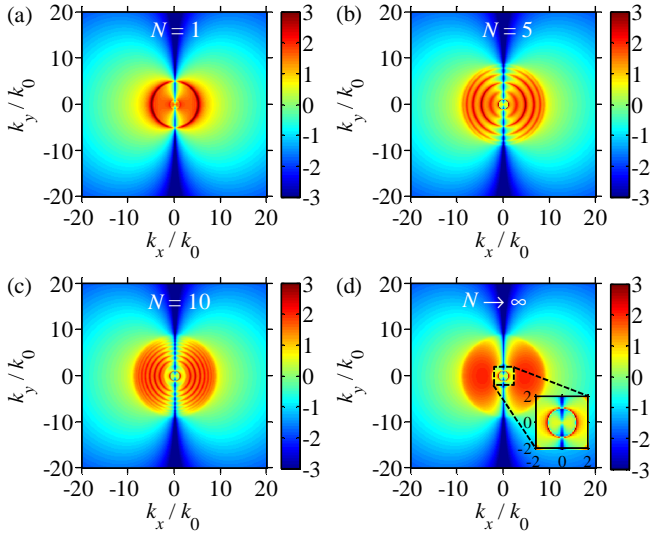

Fig. 8. Spectral power $\log_{10}\left[U_{\text{down}}^{\text{TM}} / \left(\text{Wm}^2\text{s}^2\right)\right]$ versus $k_x$ and $k_y$ emitted by the unit transverse dipole $\mathbf{p}_{00} = \hat{\mathbf{x}}$ Cm over a multilayer HM at 650 THz, for varying number of bi-layers $N$.

In Fig. 9, we report the emitted power enhancement $P / P_{\text{free space}}$ and the ratio of the power in the downward/upward direction $P_{\text{down}} / P_{\text{up}}$ as previously done in Fig. 3. The left panels are pertinent to the array of dipoles, whereas the right panels show the case of a single dipole for comparison. Various number of bi-layers $N = 1, 5, 10$ and $N \to \infty$ are analyzed. Importantly, power enhancement is observed in all cases. Furthermore, in the case of array of dipoles on top of a HM substrate with finite $N$, the spectral sampling of the spectral peaks in Fig. 8 by the FW harmonics

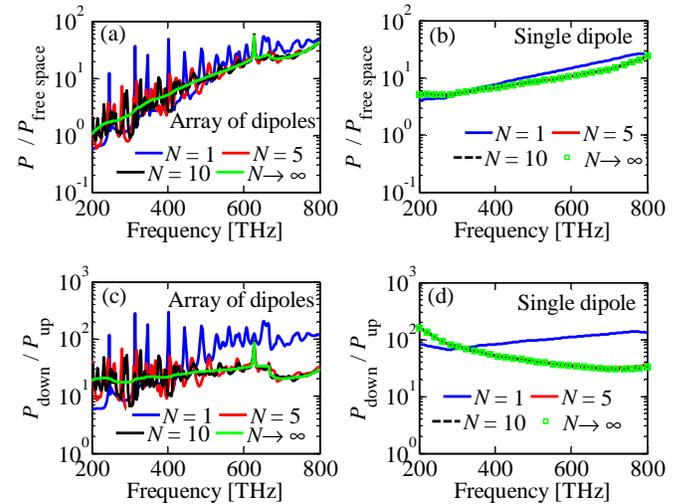

Fig. 9. Emitted power by an array of dipoles (normalized to the power emitted by the same array in free space) and ratio of power emitted towards HM and towards free space versus frequency, compared to the case of single dipole on HM. This result is calculated assuming a multilayer HM.

The power enhancement peaks for the array of dipoles can be explained by investigating the frequency evolution of the sampling points of the spectrum emitted by a single dipole. This is shown in Fig. 10 assuming $N = 5$ where $U_{\text{down}}^{\text{TM}}$ is plotted versus $k_x / k_0$ ($x$-axis) and frequency ($y$-axis), for $k_y = 0$. We also superimpose the frequency evolution of $k_{x,p} = \mathbf{k}_{t,pq} \cdot \hat{\mathbf{x}}$ normalized by $k_0$ (the $p$ index is indicated on the top of the plot) denoted by dashed lines, assuming $\mathbf{k}_{t,00} = 0.5 k_0 \hat{\mathbf{x}}$. One can note that as the frequency increases more and more FW harmonics fall in the propagating spatial spectrum of HM. At certain frequencies a $k_{x,p}$ sample coincides with a spectral peak (observed as circular spectral regions in Fig. 8) and this causes the occurrence of a narrow frequency feature in Fig. 9. Since several FW wavenumbers can meet the peaks of the power spectrum in the HM, several power emission peaks can occur when varying frequency. Therefore the finite thickness HM has very narrow frequency features in power emission which can be useful in detection applications. It is important to note that the mode represented by the peak in the region where $|k_x|$ is slightly larger than $k_0$ does not belong to the propagating spectrum of HM, but it is the long-range SPP as shown in [40] and in Appendix. The sampling of this mode by the FW with $p = -1$, $q = 0$ harmonic (pointed by the pink arrow in Fig. 10) results in the peak at 627.4 THz previously observed in Fig. 2 and Fig. 3. Note that this mode is distinct from the propagating spectrum of HM and is present even for $N \rightarrow \infty$. On the other hand, for the single dipole case the wavenumber associated to the power coupled to this SPP mode is included in the integration domain of Eq. (23) at any frequency, thus we do not observe a power peak in Fig. 2.

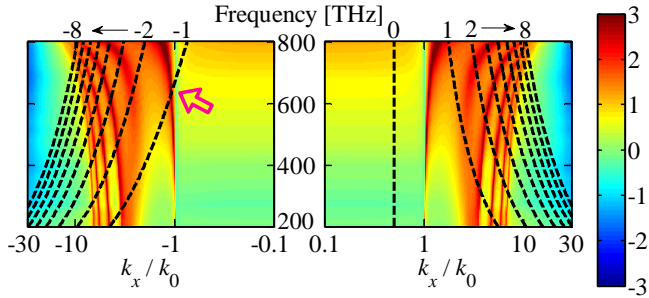

Fig. 10. Power spectrum of extraordinary TM waves carrying power in the negative $z$ direction, $\log_{10}\left[U_{\text{down}}^{\text{TM}}(\mathbf{k}_t)/(\text{Wm}^2\text{s}^2)\right]$, versus $k_x$ ($x$-axis) and frequency ($y$-axis) emitted by a a single transverse dipole with $\mathbf{p}_{00} = \hat{\mathbf{x}}$ Cm on top of a finite thickness HM with $N = 5$ layers. Black dashed curves indicate spectral sampling lines corresponding to $\mathbf{k}_{t,p0} = k_{x,p}\hat{\mathbf{x}}$ for $p = -8, -7, ..., 0, ..., 7, 8$ (denoted on top of the plot) when an array of dipoles in considered. This result is calculated assuming a multilayer HM.

## VI. EFFECT OF SOURCE SPECTRUM

In the previous sections we have considered ideal (i.e., point like) dipoles as array elements. However in general, the size and physical domain of existence of the extended electric currents of the array elements also affect the amount of power coupled to the HM. This becomes especially important when the domain of the array-element current is no more extremely subwavelength. For example if we assume an extremely flat current domain, then in Eqs. (19) and (20) the dipolar term $\mathbf{p}_{00}$ can no longer be assumed having a constant spatial spectrum. It should be replaced by $\mathbf{p}_{00}(\mathbf{k}_t)$ evaluated in terms of the sheet electric current density $\mathbf{J}_{00}(x, y)$ flowing over the unit cell area $S$, at $z = 0$, as

$$\mathbf{p}_{00}(\mathbf{k}_t) = \frac{1}{-i\omega}\iint_S \mathbf{J}_{00}(x, y) \cdot e^{-i(k_x x + k_y y)}dx dy \quad (26)$$

As an example, consider a periodic array of flat rectangular current domains with dimensions $l_x$ and $l_y$, as depicted in Fig. 11(a), on which a constant current $\mathbf{J}_{00}$ flows in the region $x \in (-l_x/2, l_x/2)$ and $y \in (-l_y/2, l_y/2)$. Then, the spectral power emitted by this array will be the FW sampling of the spectrum emitted by the isolated current sheet in Fig. 11(b), equivalent to the case described in Sec. IV for imposed dipoles. The dipolar term $\mathbf{p}_{00}(\mathbf{k}_t)$ is a function of $\mathbf{k}_t$ and is given by

$$\mathbf{p}_{00}(\mathbf{k}_t) = \frac{1}{-i\omega}\left(\mathbf{J}_{00}l_x l_y\right)\frac{\sin k_x l_x}{k_x l_x}\frac{\sin k_y l_y}{k_y l_y} \quad . \quad (27)$$

It is clear that when $k_x l_x \ll 1$ and $k_y l_y \ll 1$, then $\mathbf{p}_{00}(\mathbf{k}_t)$ becomes constant (and independent of $\mathbf{k}_t$) approaching the discrete dipole case as

$$\mathbf{p}_{00} \approx \frac{1}{-i\omega}\mathbf{J}_{00}l_x l_y. \quad (28)$$

Now let us show the impact of the source spatial spectrum on the total emitted power spectrum. For a fair comparison with the cases in Sec. III, here we will assume $\left|\mathbf{p}_{00}(\mathbf{k}_t)\right|$ having unity maximum, determined by imposing $\left|\frac{1}{-i\omega}\mathbf{J}_{00}(x, y)l_x l_y\right| = 1$ Cm.

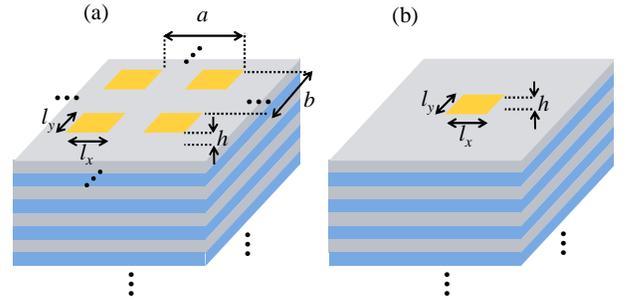

Fig. 11. The schematic of (a) a periodic array of rectangular current sheets on top of HM, and (b) a single rectangular current sheet on top of HM.

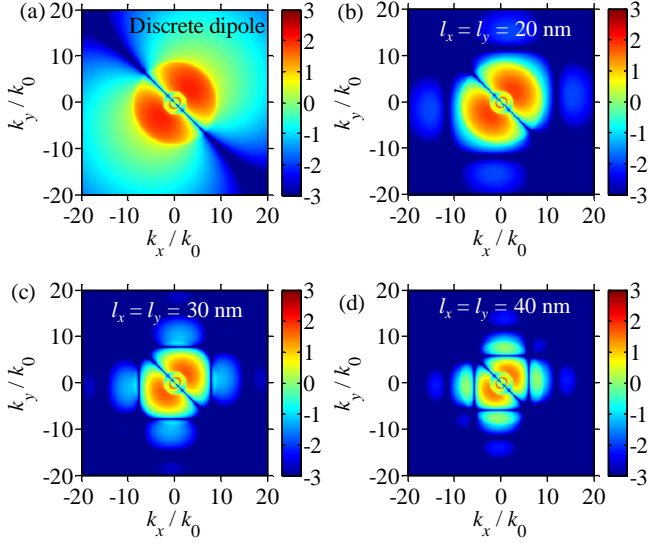

Fig. 12. Spectral power $\left(\log_{10}\left[\left(U_{\mathrm{down}}^{\mathrm{TM}}\right)/\left(\mathrm{Wm}^2\mathrm{s}^2\right)\right]\right)$ versus $k_x$ and $k_y$ emitted at 650 THz (a) by the discrete dipole $\mathbf{p}_{00}=(\hat{\mathbf{x}}+\hat{\mathbf{y}})/\sqrt{2}$ Cm and (b,c,d) by the constant sheet current $\mathbf{J}_{00}=-i\omega\mathbf{p}_{00}/\left(l_x l_y\right)$ flowing over a flat square sheet with sides $l_x=l_y=20,30,40$ nm, respectively, centered at the origin. This result is calculated assuming a multilayer HM.

In Fig. 12(a) we report the power spectrum pertaining to an ideal elementary dipole with $\mathbf{p}_{00}=(\hat{\mathbf{x}}+\hat{\mathbf{y}})/\sqrt{2}$ Cm (note that $|\mathbf{p}_{00}|=1$ Cm) and in Fig. 12(b)-(d) the ones pertaining to the sheet current $\mathbf{J}_{00}=-i\omega\mathbf{p}_{00}/\left(l_x l_y\right)$ for $l_x=l_y=20,30,40$ nm, respectively. As clearly observed, the distributed current source truncates the high $k_x$ - $k_y$ spectrum due to the double *sinc* envelope in Eq. (27) imposed by the current domain's shape and dimensions. Moreover, by visual inspection of Fig. 12, it is clear that the power spectrum with large $\mathbf{k}_t$ (which would be able to couple to the HM) is suppressed in the distributed current domain case, when compared to the discrete dipole case. This causes a decrease of the power coupled to the HM, although for electrically small current domains its impact is limited. Therefore the total power emitted by an array of current sheets (evaluated by sampling the spectrum in Fig. 12, as shown in Fig. 6) and the power emitted towards the HM will both decrease as either $l_x$ or $l_y$ increases if array periodicity is unchanged.

## VII. Conclusion

In this work, the dynamics of Floquet waves emanating from a periodic set of emitters above HM substrate is studied analytically for the first time. The power coupled to the HM from the array of dipoles is shown to encompass numerous Floquet waves with high indices. The power emitted by an array of dipolar sources (over a unit cell) is highly enhanced with respect to free-space emission; moreover this power is mostly coupled to the HM substrate. We have also observed the interesting feature that the power emitted by the array, and the one coupled to the HM, exhibit narrow frequency peaks of very strong enhancement which can be useful in detection, probing and filtering applications. The physics behind this feature associated to arrays over HM has been explained through the concept of sampling the power spatial spectrum of the isolated dipole on top of HM.

## Appendix A : Surface Plasmon Polariton Mode Supported at the Interface of Free Space and HM

We derive here the transverse wavenumber of the SPP mode associated to the free space-HM interface. We use the transverse resonance method [1] applied at the interface, i.e.,

$$Z_{\mathrm{HM}}^{\mathrm{TM}}\left(k_t\right)+Z_0^{\mathrm{TM}}\left(k_t\right)=0 \tag{29}$$

where $Z_0^{\mathrm{TM}}$ and $Z_{\mathrm{HM}}^{\mathrm{TM}}$ are evaluated using Eqs. (10) and (11), respectively for the case of homogeneous HM. (In the case a a multilayer HM, Eq. (29) is still valid but $Z_{\mathrm{HM}}^{\mathrm{TM}}$ could be evaluated using the transfer matrix method.) Then Eq. (29) leads to

$$\frac{k_{z1}}{\varepsilon_t}=-k_{z0}, \tag{30}$$

where for simplicity we have omitted the superscript TM for $k_{z1}$. Note that, when losses are neglected, a solution may exist only if both $k_{z0}$ and $k_{z1}$ are imaginary. Assuming a lossless case for simplicity, with $\varepsilon_t<0$ and $\varepsilon_z>1$ (recall that we analyzed the case with $\mathrm{Re}\{\varepsilon_t\}<0$ $\mathrm{Re}\{\varepsilon_z\}>1$ in Sec III-V), solutions of imaginary $k_{z0}$ and $k_{z1}$ may be found in the range of transverse wavenumber as

$$k_0^2<k_t^2<\varepsilon_z k_0^2. \tag{31}$$

Solving Eq. (30) for $k_t$ one has

$$\begin{aligned} &\left(k_{z1}\right)^2=\left(-\varepsilon_t k_{z0}\right)^2, \\ &k_0^2-\frac{1}{\varepsilon_z}k_t^2=\varepsilon_t\left(k_0^2-k_t^2\right) \end{aligned} \tag{32}$$

and both the left and right hand side members of the last equations should be positive. Thus,

$$\begin{aligned} &\left(\frac{1}{\varepsilon_z}-\varepsilon_t\right)k_t^2=k_0^2\left(1-\varepsilon_t\right) \\ &k_t^2=k_0^2\frac{\varepsilon_z-\varepsilon_t\varepsilon_z}{1-\varepsilon_t\varepsilon_z}. \end{aligned} \tag{33}$$

Note that the fraction term is a positive number larger than unity since $\varepsilon_z>1$, implying that $k_t^2$ satisfies the original assumption in Eq. (31). The wavenumber of the SPP wave at the interface of free space and HM thus is

$$k_t = k_0 \sqrt{\frac{\varepsilon_z - \varepsilon_t \varepsilon_z}{1 - \varepsilon_t \varepsilon_z}} \qquad (34)$$

when $\varepsilon_t < 0$ and $\varepsilon_z > 1$. For $|\varepsilon_t \varepsilon_z| \gg 1$ and $|\varepsilon_t \varepsilon_z| \gg |\varepsilon_z|$ this mode has $k_t$ slightly greater than $k_0$ which can be observed in Fig. 10 particularly at smaller frequencies. This result is in agreement with previous predictions [52, 53]. No wave solutions can be found when $\varepsilon_t < 0$ and $0 < \varepsilon_z < 1$ since $k_{z0}$ and $k_{z1}$ cannot be imaginary simultaneously. Similarly, when $\varepsilon_t > 0$ and $\varepsilon_z < 0$, $k_{z1}$ is never imaginary thus a bound mode at the surface is not supported.